# WHY NOT NOW? INTENDED TIMING IN ENTREPRENEURIAL INTENTIONS

*Antonio Rafael Ramos-Rodríguez [1] & José Aurelio Medina-Garrido [1*] & José Ruiz-Navarro [1]*

[1] *INDESS, Universidad de Cádiz, Spain*
[*] *Corresponding author, joseaurelio.medina@uca.es*

This is the submitted version accepted for publication in the "International Entrepreneurship and Management Journal". The final published version can be found at:
https://doi.org/10.1007/s11365-019-00586-5
We acknowledge that Springer holds the copyright of the final version of this work.
Please, cite this paper in this way:
Ramos-Rodríguez AR, Medina-Garrido JA, Ruiz-Navarro J (2019). Why not now? Intended timing in entrepreneurial intentions. International Entrepreneurship and Management Journal, 15:1221–1246.

**Abstract**

**Purpose** - Understanding the formation of entrepreneurial intentions is critical, given that it is the first step in the entrepreneurial process. Although entrepreneurial intention has been extensively studied, little attention has been paid on the intended timing of future entrepreneurial projects. This paper analyses entrepreneurial intentions among final-year university students after graduation in terms of the timeframe to start a business. Potentially rapid entrepreneurs and entrepreneurs-in-waiting were compared using the Theory of Planned Behaviour (TPB).

**Design/methodology/approach** - A variance-based structural equation modelling approach was used for the sample of 851 final-year university students with entrepreneurial intentions who participated in GUESSS project.

**Findings** - The results obtained contribute to the understanding of how entrepreneurial intentions are formed, particularly, how intended timing plays a moderating role in the relationships of the variables of the theoretical model of TPB. This study provides empirical evidence that significant differences exist between potential rapid entrepreneurs and entrepreneurs-in-waiting.

**Practical implications** - The findings of this study have practical implications for entrepreneurship education, and they can help policy makers develop more effective policies and programs to promote entrepreneurship.

**Originality/value** - Intention-based models have traditionally examined the intent –but not the timing– of new venture creation. However, the time elapsed between the formation of the entrepreneurial intent and the identification of a business opportunity can vary considerably. Therefore, analysing the moderating role of intended timing could be relevant to entrepreneurial intention research.

## Introduction

Entrepreneurial intentions (herein after "EI") are central to understanding entrepreneurship, since it is the first step in the process of discovering, creating, and exploiting business opportunities (Gartner et al., 1994). Exploring how EI forms is important to the understanding of entrepreneurial behaviour (Fayolle and Liñán, 2014; Shane and Venkataraman, 2000) and proves particularly valuable when the focal phenomenon is rare, obscure, or involves unpredictable time lags—a focal phenomenon such as entrepreneurship (MacMillan and Katz, 1992). The literature on entrepreneurship has made significant efforts to explain how and why new ventures are started, and valuable theoretical and empirical contributions have been made to the understanding of the early stages of the entrepreneurial process (Schlaegel and Koenig, 2014).

Although EI has been extensively studied, little attention has been paid on the intended timing of future entrepreneurial projects. Intention-based models have traditionally examined the intent –but not the timing– of new venture creation (Krueger, Reilly, and Carsrud, 2000; Tumasjan, Welpe, and Spörrle, 2013). However, the time elapsed between the formation of the entrepreneurial intent and the identification of a business opportunity can vary considerably. In addition, research on EI does not take into account inter-subject variability in the intended timing of future entrepreneurial projects, which might limit the relevance of the conclusions drawn. Therefore, analysing the moderating role of intended timing could be relevant to EI research.

There is empirical evidence of the high prevalence of EI among college students (e.g. Franke and Luthje, 2003; Lee et al., 2011; Sieger et al., 2014; Zhang et al., 2014). However, only a small percentage of students have EI in the short term and most prefer being a salaried employee first and become self-employed later –in other words, to "wait and see" before starting a business (Collins, Hannon, and Smith, 2004; Galloway and Brown, 2002; Sieger et al., 2014). This may be due to the fact that students initially feel they lack the resources and essential skills, knowledge, and experiences necessary to start a business and that these can be best gained through employment than through self-employment (Collins et al., 2004).

The purpose of this research study is to investigate the intended timing of future entrepreneurial activities among final-year college students. This paper analyses the entrepreneurial intentions of college students, with particular focus on the moderating role of the timeframe within which students intend to become entrepreneurs. The research model is adapted from the model of entrepreneurial intentions developed by Liñán and Chen (2009) and is used to investigate whether the relationships between the variables of the model are affected by entrepreneurial timing.

**Theoretical framework and hypotheses**

EI refers to the target behaviour of starting a business (Krueger et al., 2000). Intentions are the best predictor of planned behaviour and an unbiased predictor of action, even where time lags exist (Bagozzi et al., 1989). Thus, intention models are ideal for assessing entrepreneurship behaviour (Bird, 1988; Katz and Gartner, 1988). Among the existing EI models, Ajzen's (1991, 2002) Theory of Planned Behaviour (TPB) is the predominant model (Fayolle and Liñán, 2014). The TPB analyses if the attitude toward the behaviour (hereafter ATB), subjective norm (also known as social norm; hereafter "SN"), and perceived behavioural control (hereafter "PBC") are positively related to one's intention to perform a certain behaviour. According to the TPB, intention indicates the effort that a person is willing to make to perform a behaviour. Consequently, the higher the EI, the higher the effort that a person will make to perform the entrepreneurial behaviour.

Having an EI does not mean that the firm will be created shortly. It may be either a short- or long-term process (Shook et al., 2003). Works analysing the time-lag between the formation of an EI and entrepreneurial behaviour are still scarce (Fayolle and Liñán, 2014; Kwong and Thompson, 2016). Nevertheless, the reasons why some entrepreneurs are able to create a new firm faster than others are still unclear.

Though there has been some interest in entrepreneurial timing (Bird, 1992; Bird and West, 1997; Capelleras et al., 2010), the existing literature on entrepreneurship rarely distinguishes between potential entrepreneurs who are starting a business immediately and "wait-and-see" entrepreneurs. Whereas only a few business school graduates immediately start a venture, most graduates prefer working for others before (Collins et al., 2004; Galloway and Brown, 2002). The intend to wait can partly be attributed to a perceived lack of some of the resources, skills, knowledge, experiences, or contacts required to start a business (Kwong and Thompson, 2016). These facts justify the need to investigate whether the intended timing of venture creation influences the formation of EI in potential rapid entrepreneurs (PRE) as compared to entrepreneurs-in-waiting (EIW) (Kwong and Thompson, 2016). Figure 1 summarizes the conceptual model and hypotheses.

Figure 1. Research model and hypotheses

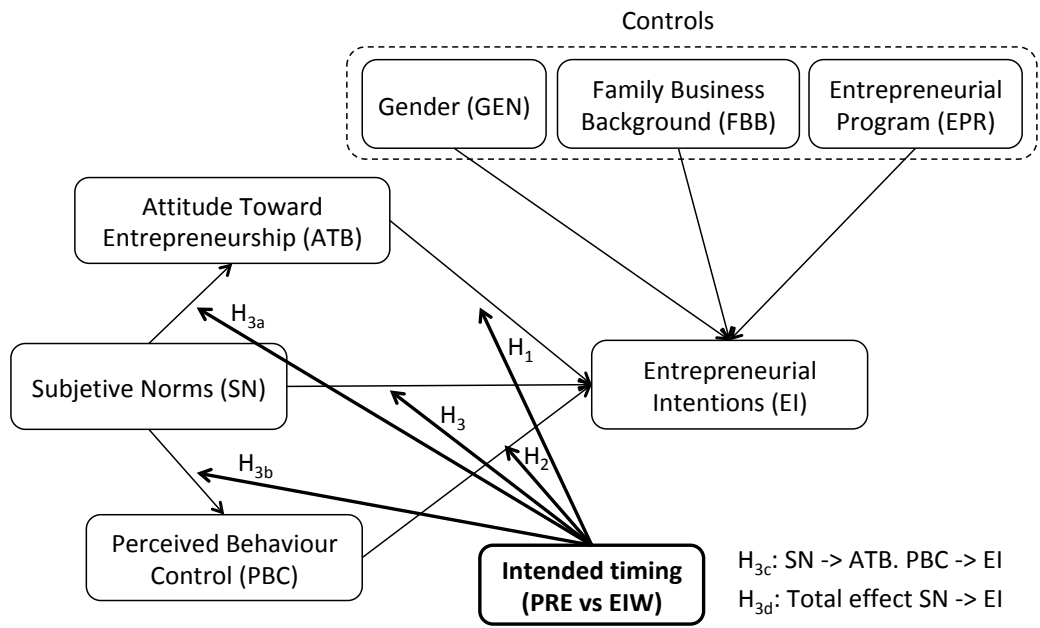

*Differences between PRE and EIW in the ATB-EI relationship*

ATB refers to the degree to which a person has a favourable or unfavourable evaluation of a certain behaviour (Ajzen, 1991, 2002). According to the TPB, ATB is positively related to an individual's intention to perform the behaviour (Ajzen, 1991, 2002). Thus, the attitude toward entrepreneurship indicates the degree to which the individual holds a positive or negative appraisal of becoming an entrepreneur (Liñán and Chen, 2009), and this attitude should be positively related to EI (Autio, Keeley, Klofsten, Parker, and Hay, 2001). People develop attitudes from their beliefs about the consequences of performing a certain behaviour (Ajzen, 2005). Thus, the attitude toward entrepreneurship represents an individual's appraisal of the possible outcomes of setting up a new firm and their attractiveness (Heuer and Kolvereid, 2014). The perception of favourable outcomes of entrepreneurship could stimulate EI (Volery et al., 2013). The outcomes of entrepreneurship include intrinsic and extrinsic rewards such as autonomy, authority, financial rewards, self-realization (Douglas and Fitzsimmons, 2013; Fretschner and Weber, 2013; Kolvereid and Isaksen, 2006; Saeed et al.,

2015; Vanevenhoven and Liguori, 2013), or self-employment (Segal et al., 2005). On the one hand, and in accordance with Kwong and Thompson (2016), the appraisal of the outcomes of entrepreneurship of PRE may differ from that of EIW and, therefore, their attitude toward entrepreneurship may vary. Thus, PRE wish to be leaders and have authority over others, dislike working for others, display a greater willingness to take on additional responsibilities, and they can withstand greater stress or cope with it to a greater extent than EIW (Bluedorn and Martin, 2008; Kwong and Thompson, 2016). On the other hand, the relationship between EIW's attitude toward entrepreneurship and EI may be conditioned by obtaining essential skills, experience, and finance; otherwise their EI may tend to dissipate (Carsrud and Brännback, 2011; Kwong and Thompson, 2016). For these reasons, PRE have a more favourable appraisal of entrepreneurial behaviour than entrepreneurs-in-waiting (EIW), and their attitude toward the behaviour (ATB) may have a stronger positive effect on EI than in EIW.

The arguments exposed above led to the development of the following hypothesis:

> H1. ATB has a stronger positive effect on EI in PRE than in EIW.

*Differences between PRE and EIW in the PBC-EI relationship*

According to the TPB, perceived behavioural control (PBC) refers to an individual's perception of the ease or difficulty in performing certain behaviour (Ajzen, 1991). In the field of entrepreneurship, the concept of PBC is similar to those of self-efficacy (Bandura, 1997) and perceived feasibility (Shapero and Sokol, 1982). All three concepts refer to the perception of one's capacity to start a business (Liñán and Chen, 2009). PBC may be determined by internal or external factors that can facilitate or impede the performance of a behaviour (Ajzen and Cote, 2008). These factors include the availability of resources, opportunities, information, skills, emotions, dependence on others, and own or others' experience in the past (Ajzen, 2005, 2011, 2012).

Ajzen (1991) states that the greater the PBC, the harder the effort made to perform the intended behaviour. Consequently, the greater the degree to which the behaviour toward starting a business can be controlled, the greater the intensity of EI. According to Autio et al. (2001), among other TPB determinants, PBC has an important influence on EI.

Research on the impact of PBC on entrepreneurship suggests that a person's PBC may determine the time lag between the formation of an EI and the creation of a new firm (Kwong and Thompson, 2016). Thus the relationship between PBC and EI could vary depending on the type of potential entrepreneur. Regarding this relationship between PBC and EI, Kwong and Thompson (2016) found no evidence that PRE feel more confident that they have the skills and knowledge required to start a business than EIW. Yet, EIW may decide not to create their firm immediately because of a perceived lack of some essential skills, knowledge, and experiences. By taking special training (Carter and Collinson, 1999) and working for others first (Collins et al., 2004), EIW ensure that they will have control over the entrepreneurial situation in the future. Thus, EIW's expectations of a high level of PBC in the near future cause them to have a high level of IE in the long term. Therefore, the relationship between PBC and EI may be weak in PRE.

The arguments exposed above led to the development of the following hypothesis:

> H2. PBC has a stronger positive effect on EI in EIW than in PRE.

*Differences between PRE and EIW in the SN-EI relationship*

Social norm (SN) refers to the perceived social pressure to perform or not a certain behaviour (Ajzen, 1991). SN influences an individual's beliefs on significant others' approval or disapproval of the behaviour (Ajzen, 2005). Significant others may include a person's parents, spouse, close friends, co-workers and even experts in the behaviour of interest (Malebana, 2014).

SN influences on entrepreneurial behaviour could determine the decision to perform or not entrepreneurial behaviours (Liñán and Chen, 2009). Individuals are more likely to form EI when significant others approve the entrepreneurial behaviour or are entrepreneurs themselves (Laspita, Breugst, Heblich, and Patzelt, 2012; Pablo-Lerchundi, Morales-Alonso, and González-Tirados, 2015). The influence of SN on EI is supported by the literature on entrepreneurship (Engle et al., 2010; Gird and Bagraim, 2008; Iakovleva, Kolvereid, and Stephan, 2011; Souitaris, Zerbinati, and Al-Laham, 2007).

Significant others' support is also essential in determining the speed of venture creation. Kwong and Thompson (2016) observed that PRE may be more likely to feel that they are supported by significant others than EIW. The EI of EIW is less likely to be influenced by support from significant others because support for future entrepreneurial activity may be weak or vague. On the contrary, support from PRE's significant others may strengthen their EI, as the immediacy of entrepreneurial behaviour requires an actual and stronger commitment with them. Consequently, it would be interesting to analyse if the influence of perceived support on EI varies with the type of potential entrepreneur (PRE versus EIW).

The following hypothesis was formulated:

>  H3. SN has a stronger positive effect on EI in PRE than in EIW.

On the other hand, the literature has found that SN has a minor role in the TPB model. However, this statement is not accurate in the area of entrepreneurship (Liñán and Chen, 2009). The fact that SN has not been found to be significant (Autio et al., 2001; Krueger et al., 2000) might be due to the indirect effects of SN on EI (Liñán and Chen, 2009). There is evidence that SN influences not only EI but also ATB and PBC (Allen, 2000; Kwong and Thompson, 2016; Schlaegel and Koenig, 2014). Thus, the direct effects of SN on ATB and PBC may have an indirect impact on EI.

Support from significant others has a positive effect on individual perceptions regarding one's capability to accomplish the entrepreneurial behaviour (perception of capability) and its attractiveness (perception of attractiveness) (García-Rodríguez, Gil-Soto, Ruiz-Rosa, and Sene, 2015; Liñán, Rodríguez-Cohard, and Rueda-Cantuche, 2011). The concept of perceived capability assumes that SN has an impact on PBC, whereas the concept of perceived attractiveness assumes an influence of SN on ATB. Individuals are more likely to perceive the entrepreneurial behaviour as something favourable when it is approved and positively valued by significant others (Liñán, Nabi, and Krueger, 2013). On the other hand, the influence of SN on PBC arises when potential entrepreneurs consider the resources and emotional support that significant others may make available to them (Allen, 2000; Kwong and Thompson, 2016).

The direct effects of SN on ATB and PBC, and their likely indirect and total effects (direct plus indirect) on EI may differ between PRE and EIW. As stated above, significant others' support is not as clear in EIW as in PRE, since their EI may be perceived as probable in EIW, but not as much as it is in PRE. PRE who are supported by their significant others would have a stronger perception of their own entrepreneurial capability, and entrepreneurship would be a more appealing idea (García-Rodríguez, Gil-Soto, Ruiz-Rosa, and Sene, 2015; Liñán, Rodríguez-Cohard, and Rueda-Cantuche, 2011) . The former has a positive impact on PBC, whereas the latter has a positive influence in ATB.

The above arguments led to the development of the following hypotheses about the direct,effects of SN on ATB and PBC:

> H3a. SN has a stronger positive effect on ATB in PRE than in EIW.
>
> H3b. SN has a stronger positive effect on PBC in PRE than in EIW.

Finally, the following hypotheses about the indirect and total effects of SN on EI are proposed:

> H3c. The mediating effect of ATB and PBC on the relationship between SN and EI will be stronger in PRE than in EIW.
>
> H3d. The total effect of SN on EI will be stronger in PRE than in EIW.

**Research methodology**

*Sample and data collection*

Data were extracted from the Spanish version of the 2013-2014 Global University Entrepreneurial Spirit Students' Survey (GUESSS). GUESSS is an international research project based on the geographical and temporal comparison of EI and entrepreneurial activities of students. This project provides an in-depth analysis of this dataset, shedding a nuanced light on students' entrepreneurial intentions and concrete activities (Sieger et al., 2014).

The 2013-2014 GUESSS edition investigated and compared entrepreneurial attitudes, intentions, and activities of students at more than 700 universities from 34 countries between October 2013 and March 2014. In total, 21 Spanish universities participated in GUESSS. Data were gathered by means of an online survey. The students at these universities received an e-mail with a neutrally worded invitation to participate in a study on their 'future career choice' and a link to the online survey.

In Spain, the number of students addressed was 126,870 and the valid response rate was 8.31%. Therefore, the sample size was 10,545. To rely on a more homogeneous sample, only final-year students were considered. This approach, despite its known limitations, is regarded as adequate for investigating EI (Krueger et al., 2000) and is consistent with the study design applied in previous studies on entrepreneurial intentions (Autio et al., 2001; Franke and Luthje, 2003; Kristiansen and Indarti, 2004; Krueger et al., 2000). Only final-year students were included in our study because making a career choice was a real scenario. Also, as a group, final-year students present a broad spectrum of intentions and attitudes towards entrepreneurship (Fatoki, 2014a, 2014b; Gird and Bagraim, 2008; Krueger et al., 2000; Wurthmann, 2014). After the selection criteria were applied, the sample size was reduced to 851 subjects.

A central question in GUESSS was what students intended to do in the short term and within five years after completing their studies. The reason for using the five-year time frame is that entrepreneurs often work as salaried employees before starting their own business (Krueger et al., 2000). This time frame allowed us to differentiate potential rapid entrepreneurs from entrepreneurs-in-waiting. "Rapid entrepreneurs" were defined as students who intended to start a business immediately after graduation, whereas "entrepreneurs-in-waiting" were defined as those who intended to wait five years before going into business. Based on these criteria, 121 potential rapid entrepreneurs and 730 potential "wait-and-see" entrepreneurs were identified. Table 1 shows descriptive statistics for the whole sample and by groups.

Table 1. Descriptive statistics of respondents' characteristics

|  | All | EIW | PRE |
|---|---|---|---|
| Age |  |  |  |
|   Mean | 24.08 | 23.86 | 25.24 |
|   Std. Deviation | 3.50 | 3.30 | 4.49 |
| Gender |  |  |  |
|   Male | 51.42% | 48.90% | 66.12% |
| Field of study |  |  |  |
|   Arts and Humanities | 11.48% | 11.06% | 13.91% |
|   Science | 3.48% | 3.03% | 6.09% |
|   Health Sciences | 8.00% | 8.64% | 4.35% |
|   Social and Legal Sciences | 46.58% | 46.21% | 48.70% |
|   Engineering and Architecture | 30.45% | 31.06% | 26.96% |

*Measurement instruments*

The dependent variable –EI– was measured using the 6-item scale proposed by Liñán and Chen (2009). Respondents were asked six questions and had to answer on a 7-point Likert scale (1 = strongly disagree, 7 = strongly agree), e.g. "I am ready to do anything to be an entrepreneur". "My professional goal is to become an entrepreneur." "I will make every effort to start and run my own firm." "I am determined to create a firm in the future." "I have very seriously thought of starting a business." "I have the strong intention to start a business someday".

ATB was measured using the aggregate attitude scale adapted from Liñán and Chen (2009) with the measures of Ajzen (1991, 2001), Fayolle et al. (2006), Kolvereid and Isaksen (2006), and Krueger et al. (2000). The students were asked five questions to be answered on a 7-point Likert scale (1 = strongly disagree, 7 = strongly agree), e.g. "Being an entrepreneur implies more advantages than disadvantages to me." "A career as entrepreneur is attractive for me." "If I had the opportunity and resources, I would become an entrepreneur." "Being an entrepreneur would entail great satisfactions for me." "Among various options, I would rather become an entrepreneur."

SN captures the reaction that individuals expect from close peers if a certain behaviour is executed. SN was measured using a scale proposed by Liñán and Chen (2009). Students were asked about the potential reactions of people in their environment if they (the students) pursued a career as an entrepreneur. People in their environment (or groups of people) included close family members, friends, and fellow students. Responses were anchored at 1 (very negatively) and 7 (very positively).

PBC was measured using the scale proposed by Souitaris et al., (2007). Students were asked whether they agreed with the statement (on a scale from 1 — strongly disagree to 7 — strongly agree): "For me, being self-employed would be very easy." "If I wanted to, I could easily pursue a career as self-employed." "As self-employed, I would have complete control over the situation." "If I become self-employed, the chances of success would be very high".

Analysis was performed after controlling for students' gender (GEN), family business background (FBB), specific entrepreneurship education, and attendance to an entrepreneurship program (EPR), to account for known or potential confounding factors in any of the constructs of the model. To control for gender, a dummy variable ("0" for male, "1" for female) was employed. To control for family business background, a corresponding dummy variable ("0" if parents or family members like siblings, grandparents, etc. were not self-employed, and "1" if they were) was applied. Regarding entrepreneurship education, a dummy variable was used and coded "0" if the student had not attended any course on entrepreneurship so far, had not attended at least an elective or compulsory entrepreneurship course as part of their studies, or had not been engaged in a specific program on entrepreneurship; and "1" if they did. There is abundant empirical evidence of the relationship between these three variables and entrepreneurial intentions (see, for example, Bae et al., 2014; Camelo-Ordaz et al., 2016; Zellweger et al., 2011)

*Data analysis*

Hypothesis testing was performed using a variance-based structural equation modelling approach (PLS-SEM) with Smart PLS 3.1.9 software (Ringle et al., 2015). Path coefficients were estimated for the total sample and for each group or subsample. For this purpose, responses were divided into two groups according to entrepreneurial timing (group 1 = rapid entrepreneurs; group 2 = entrepreneurs-in-waiting). Finally, differences between path coefficients were analysed. Significance of differences was determined by Partial Least Squares Multi-Group Analysis (PLS-MGA). This method is a non-parametric significance test for assessing differences in results between groups, which builds on PLS-SEM bootstrapping results. A result was considered significant at the 5% level and if the *p*-value was <0.05 or >0.95 for a certain difference between groups in path coefficients (Sarstedt et al., 2011).

**Results**

In accordance with recommendations of Chin (2010) and Hair et al. (2014), the results obtained were analysed first by evaluating the statistical model and then the statistical significance of model parameters. This process ensures the reliability and validity of measurements before any conclusions are drawn about correlations between constructs.

*Evaluation of the measurement model*

Table 2 shows the results of the reliability analysis. All the indicators of the measurement model were maintained, although some of them did not meet the minimum required threshold ($\lambda \geq 0.707$). This decision was based on the fact that the AVE exceeded 0.5 for all latent variables. This means that eliminating these indicators in an attempt to attain the minimum AVE threshold would be pointless (Hair et al., 2014). Therefore, all indicators were maintained because they contributed to content validity.

To evaluate scale reliability, composite reliability (Table 2) was estimated. (Nunally, 1978) proposed using 0.7 for a modest reliability and a stricter 0.8 for basic research. As Table 3 shows, all constructs exceeded these values comfortably.

Table 2. Measurement model

| CONSTRUCTS /INDICATORS | Total sample; n=851 | | | PRE; n=121 | | | EIW; n=730 | | |
|---|---|---|---|---|---|---|---|---|---|
| | Loadings | Composite Reliability | Average Variance Extracted (AVE) | Loadings | Composite Reliability | Average Variance Extracted (AVE) | Loadings | Composite Reliability | Average Variance Extracted (AVE) |
| EI | | 0.928 | 0.685 | | 0.927 | 0.680 | | 0.927 | 0.681 |
| ei1 | 0.718 | | | 0.639 | | | 0.725 | | |
| ei2 | 0.835 | | | 0.848 | | | 0.832 | | |
| ei3 | 0.871 | | | 0.810 | | | 0.878 | | |
| ei4 | 0.907 | | | 0.893 | | | 0.908 | | |
| ei5 | 0.767 | | | 0.849 | | | 0.751 | | |
| ei6 | 0.851 | | | 0.882 | | | 0.842 | | |
| ATB | | 0.906 | 0.660 | | 0.906 | 0.661 | | 0.905 | 0.658 |
| atb1 | 0.655 | | | 0.647 | | | 0.654 | | |
| atb2 | 0.848 | | | 0.831 | | | 0.849 | | |
| atb3 | 0.788 | | | 0.754 | | | 0.794 | | |
| atb4 | 0.881 | | | 0.927 | | | 0.874 | | |
| atb5 | 0.868 | | | 0.877 | | | 0.865 | | |
| PBC | | 0.845 | 0.578 | | 0.850 | 0.588 | | 0.843 | 0.573 |
| pbc1 | 0.794 | | | 0.827 | | | 0.787 | | |
| pbc2 | 0.724 | | | 0.712 | | | 0.717 | | |
| pbc3 | 0.714 | | | 0.703 | | | 0.718 | | |
| pbc4 | 0.804 | | | 0.817 | | | 0.802 | | |
| SN | | 0.835 | 0.631 | | 0.867 | 0.685 | | 0.824 | 0.615 |
| sn1 | 0.863 | | | 0.812 | | | 0.885 | | |
| sn2 | 0.850 | | | 0.901 | | | 0.831 | | |
| sn3 | 0.653 | | | 0.764 | | | 0.610 | | |

Convergent validity is evaluated by AVE. Fornell and Larcker (1981) recommend that AVE exceeds 0.5, which would mean that more than 50% of the construct's variance is due to its indicators rather than to the indicators of the rest of constructs. As Table 2 shows, all constructs had AVE values exceeding 0.5. For the single-item construct (gender, family business background and entrepreneurship program in the model), CR and AVE were found not to be appropriate measures since the indicator's outer loading was fixed at 1.00.

Table 3. Discriminant validity

|  |  | Total sample; n=851 ||||||| Potential rapid entrepreneurs (PRE); n=121 ||||||| Entrepreneurs-in-waiting- (EIW); n=730 |||||||
|---|---|---|---|---|---|---|---|---|---|---|---|---|---|---|---|---|---|---|---|---|---|---|
|  |  | ATB | EI | EPR | FBB | GEN | PBC | SN | ATB | EI | EPR | FBB | GEN | PBC | SN | ATB | EI | EPR | FBB | GEN | PBC | SN |
| Fornell-Larcker | ATB | **0.812** |  |  |  |  |  |  | **0.813** |  |  |  |  |  |  | **0.811** |  |  |  |  |  |  |
|  | EI | 0.729 | **0.827** |  |  |  |  |  | 0.800 | **0.825** |  |  |  |  |  | 0.717 | **0.825** |  |  |  |  |  |
|  | EPR | 0.062 | 0.080 | **1.000** |  |  |  |  | -0.064 | -0.026 | **1.000** |  |  |  |  | 0.073 | 0.081 | **1.000** |  |  |  |  |
|  | FBB | 0.058 | 0.045 | 0.052 | **1.000** |  |  |  | 0.038 | 0.140 | 0.238 | **1.000** |  |  |  | 0.061 | 0.031 | 0.023 | **1.000** |  |  |  |
|  | GEN | -0.034 | -0.029 | -0.048 | 0.020 | **1.000** |  |  | -0.029 | 0.046 | -0.041 | -0.025 | **1.000** |  |  | -0.021 | -0.017 | -0.038 | 0.027 | **1.000** |  |  |
|  | PBC | 0.387 | 0.408 | 0.026 | 0.086 | -0.044 | **0.760** |  | 0.274 | 0.250 | -0.050 | -0.006 | -0.067 | **0.767** |  | 0.401 | 0.425 | 0.028 | 0.104 | -0.027 | **0.757** |  |
|  | SN | 0.218 | 0.170 | 0.005 | 0.090 | 0.108 | 0.206 | **0.794** | 0.394 | 0.304 | -0.017 | 0.127 | 0.056 | 0.228 | **0.828** | 0.196 | 0.162 | 0.011 | 0.085 | 0.120 | 0.211 | **0.784** |
| HTMT | ATB |  |  |  |  |  |  |  |  |  |  |  |  |  |  |  |  |  |  |  |  |  |
|  | EI | 0.82 |  |  |  |  |  |  | 0.895 |  |  |  |  |  |  | 0.807 |  |  |  |  |  |  |
|  | EPR | 0.064 | 0.083 |  |  |  |  |  | 0.077 | 0.065 |  |  |  |  |  | 0.077 | 0.085 |  |  |  |  |  |
|  | FBB | 0.063 | 0.047 | 0.052 |  |  |  |  | 0.082 | 0.164 | 0.238 |  |  |  |  | 0.067 | 0.033 | 0.023 |  |  |  |  |
|  | GEN | 0.038 | 0.051 | 0.048 | 0.02 |  |  |  | 0.047 | 0.068 | 0.041 | 0.025 |  |  |  | 0.026 | 0.056 | 0.038 | 0.027 |  |  |  |
|  | PBC | 0.453 | 0.478 | 0.037 | 0.104 | 0.076 |  |  | 0.315 | 0.288 | 0.095 | 0.125 | 0.09 |  |  | 0.471 | 0.497 | 0.052 | 0.124 | 0.087 |  |  |
|  | SN | 0.261 | 0.183 | 0.04 | 0.091 | 0.126 | 0.248 |  | 0.478 | 0.353 | 0.075 | 0.144 | 0.127 | 0.265 |  | 0.227 | 0.164 | 0.032 | 0.081 | 0.135 | 0.249 |  |

Henseler et al. (2016) recommend using the HTMT criterion to assess discriminant validity and suggest that the HTMT value should be below 0.90. As shown in Table 3, the square root of the AVE for all the latent variables of the model was greater than all correlations between variables, which means that all constructs were more strongly related to their own indicators than to those of the other constructs. On the other hand, all values of the HTMT criteria were below 0.9.

In summary, according to PLS analyses, the measurement model was completely satisfactory for the model, both with the whole sample and with each subsample (rapid and in-waiting entrepreneurs).

*Evaluation of the structural model*

Table 4 shows path coefficients, indirect effects, total effects, t-values, variance explained ($R^2$) in the endogenous constructs, and the cross-validated redundancy index ($Q^2$). Relationships can be determined in both samples by examining the sign and statistical significance of t-values. The evaluation of significance of differences in path coefficients was carried out using the bootstrapping resampling technique. Subsamples were created following the recommendations of Hair et al. (2014) using 5,000 observations randomly extracted and with replacement of the original data set. This was a sufficiently large number to ensure that results were stable. Apart from the above and following the recommendations of the literature, an additional evaluation with a non-parametric test was performed to calculate confidence intervals. To evaluate possible indirect and total effects between the latent variables the previous procedure was repeated to evaluate these relations (see Table 4).

A measure of the predictive power of a model is the $R^2$ value of dependent latent variables, which indicates the share of the construct's variance explained by the model. Chin (1998) recommends the following thresholds: from 0.67, "substantial"; from 0.33, "moderate"; and from 0.19, "weak". The main endogenous constructs (EI) attained $R^2$ values between 0.552 and 0.659, which are considered moderate values (Chin, 2010). The subsample of potential rapid entrepreneur reached a $R^2$ value near substantial.

Table 4. Structural model results for total, potential rapid entrepreneurs and in-waiting-entrepreneurs samples. Confidence intervals (Bias-correct 95% confidence intervals)

| Path Coefficients | Total sample; n=851 | | | | | Potential rapid entrepreneurs (PRE); n=121 | | | | | Entrepreneurs-in-waiting- (EIW); n=730 | | | | |
|---|---|---|---|---|---|---|---|---|---|---|---|---|---|---|---|
| | β | 5.0% | 95.0% | t-value (bootstrap) | Explained variance | β | 5.0% | 95.0% | t-value (bootstrap) | Explained variance | β | 5.0% | 95.0% | t-value (bootstrap) | Explained variance |
| **EI** | ($R^2$=0.552; $Q^2$=0.374) | | | | | ($R^2$=0.659; $Q^2$=0.429) | | | | | ($R^2$=0.538; $Q^2$=0.365) | | | | |
| ATB ->EI | 0.672 *** | 0.634 | 0.708 | 29.824 | 48.99% | 0.801 *** | 0.699 | 0.894 | 13.665 | 64.08% | 0.649 *** | 0.609 | 0.690 | 26.119 | 46.53% |
| PBC ->EI | 0.150 *** | 0.105 | 0.195 | 5.495 | 6.12% | 0.046 n.s. | -0.050 | 0.146 | 0.766 | 1.15% | 0.167 *** | 0.118 | 0.218 | 5.456 | 7.10% |
| SN ->EI | -0.007 n.s. | -0.048 | 0.034 | 0.301 | 0.12% | -0.042 n.s. | -0.147 | 0.055 | 0.680 | 1.28% | 0.001 n.s. | -0.046 | 0.049 | 0.041 | 0.02% |
| EPR ->EI | 0.035 n.s. | -0.003 | 0.073 | 1.493 | 0.28% | 0.003 n.s. | -0.098 | 0.095 | 0.044 | 0.01% | 0.029 n.s. | -0.013 | 0.073 | 1.121 | 0.23% |
| FBB ->EI | -0.008 n.s. | -0.047 | 0.030 | 0.349 | 0.04% | 0.116 * | 0.013 | 0.231 | 1.745 | 1.62% | -0.027 n.s. | -0.069 | 0.015 | 1.039 | 0.08% |
| GEN ->EI | 0.003 n.s. | -0.034 | 0.042 | 0.139 | 0.01% | 0.078 n.s. | -0.010 | 0.164 | 1.479 | 0.36% | 0.003 n.s. | -0.040 | 0.046 | 0.125 | 0.01% |
| **ATB** | ($R^2$=0.048; $Q^2$=0.031) | | | | | ($R^2$=0.155; $Q^2$=0.095) | | | | | ($R^2$=0.038; $Q^2$=0.025) | | | | |
| SN -> ATB | 0.218 *** | 0.164 | 0.280 | 6.175 | | 0.394 *** | 0.244 | 0.555 | 4.190 | | 0.196 *** | 0.141 | 0.262 | 5.332 | |
| **PBC** | ($R^2$=0.042; $Q^2$=0.022) | | | | | ($R^2$=0.052; $Q^2$=0.021) | | | | | ($R^2$=0.045; $Q^2$=0.023) | | | | |
| SN ->PBC | 0.206 *** | 0.157 | 0.264 | 6.412 | | 0.228 ** | 0.118 | 0.373 | 2.707 | | 0.211 *** | 0.159 | 0.274 | 6.047 | |
| **Indirect Effects** | | | | | | | | | | | | | | | |
| SN -> ATB-PBC->EI | 0.177 *** | 0.138 | 0.223 | 6.815 | | 0.326 *** | 0.195 | 0.477 | 3.806 | | 0.162 *** | 0.124 | 0.210 | 6.189 | |
| **Total Effects** | | | | | | | | | | | | | | | |
| SN ->EI | 0.170 *** | 0.121 | 0.226 | 5.389 | | 0.285 ** | 0.122 | 0.454 | 2.796 | | 0.164 *** | 0.115 | 0.223 | 4.941 | |

*** $p< 0.001$. (based on t(4999). one-tailed test); t(0.05. 4999) = 1.645. t(0.01. 4999) = 2.327. t(0.001. 4999) = 3.092; Bootstrapping based on n = 5000 subsamples; ns = not significant

The predictive relevance of the theoretical/structural model was assessed using the cross-validated redundancy index ($Q^2$) for endogenous constructs. The predictive relevance of the model proposed was confirmed by the fact that all $Q^2$ values were greater than 0 (Chin, 2010). Table 4 shows the amount of variance that each of the dependent variables account for in each dependent variable. The greater value was obtained for ATB in the rapid entrepreneur subsample when analysing the explained variance in EI (64.08%). For the whole sample, ATB also explained most EI variance (48.99%).

A primary concern when comparing path coefficients across groups is ensuring measurement invariance, i.e. that the same construct was being measured across groups (Hair et al., 2014). Table 5 shows the results of measurement invariance tests. The proposed measurement model did not vary significantly when the intended timing of future entrepreneurial activities was taken into account. In other words, factor loadings for the same indicators were virtually invariant between rapid and in-waiting entrepreneurs, which confirms metric invariance (Afonso et al., 2012).

Table 5. Metric invariance testing. Multigroup analysis

|  | | PLS-MGA | Parametric Test | | Welch-Satterthwait Test | |
|---|---|---|---|---|---|---|
|  | Outer Loadings - diff (\|PRE-EIW\|) | p-value (PRE vs EIW) | t-value (PRE vs EIW) | p-value (PRE vs EIW) | t-value (PRE vs EIW) | p-value (PRE vs EIW) |
| EPR | 0.000 | 0.681 n.s. | 0.112 | 0.911 n.s. | 0.270 | 0.787 n.s. |
| FBB | 0.000 | 0.297 n.s. | 0.204 | 0.839 n.s. | 0.491 | 0.623 n.s. |
| GEN | 0.000 | 0.216 n.s. | 0.263 | 0.793 n.s. | 0.639 | 0.523 n.s. |
| ei1 | 0.087 | 0.896 n.s. | 1.422 | 0.156 n.s. | 1.219 | 0.225 n.s. |
| ei2 | 0.016 | 0.321 n.s. | 0.396 | 0.692 n.s. | 0.409 | 0.683 n.s. |
| ei3 | 0.067 | 0.940 n.s. | 2.222 | 0.027 * | 1.366 | 0.174 n.s. |
| ei4 | 0.015 | 0.667 n.s. | 0.673 | 0.501 n.s. | 0.501 | 0.617 n.s. |
| ei5 | 0.097 | 0.016 * | 1.889 | 0.059 n.s. | 2.389 | 0.018 * |
| ei6 | 0.040 | 0.110 n.s. | 1.103 | 0.270 n.s. | 1.238 | 0.218 n.s. |
| atb1 | 0.007 | 0.505 n.s. | 0.091 | 0.927 n.s. | 0.086 | 0.932 n.s. |
| atb2 | 0.018 | 0.582 n.s. | 0.435 | 0.664 n.s. | 0.339 | 0.735 n.s. |
| atb3 | 0.041 | 0.708 n.s. | 0.650 | 0.516 n.s. | 0.603 | 0.548 n.s. |
| atb4 | 0.053 | 0.010 * | 1.855 | 0.064 n.s. | 2.631 | 0.009 ** |
| atb5 | 0.012 | 0.322 n.s. | 0.361 | 0.719 n.s. | 0.344 | 0.731 n.s. |
| pbc1 | 0.040 | 0.224 n.s. | 0.539 | 0.590 n.s. | 0.427 | 0.670 n.s. |
| pbc2 | 0.006 | 0.434 n.s. | 0.059 | 0.953 n.s. | 0.041 | 0.967 n.s. |
| pbc3 | 0.015 | 0.519 n.s. | 0.179 | 0.858 n.s. | 0.143 | 0.887 n.s. |
| pbc4 | 0.016 | 0.352 n.s. | 0.286 | 0.775 n.s. | 0.217 | 0.829 n.s. |
| sn1 | 0.073 | 0.870 n.s. | 0.719 | 0.472 n.s. | 1.067 | 0.288 n.s. |
| sn2 | 0.070 | 0.070 n.s. | 0.679 | 0.497 n.s. | 1.314 | 0.190 n.s. |
| sn3 | 0.154 | 0.093 n.s. | 0.729 | 0.466 n.s. | 1.273 | 0.205 n.s. |

n.s. = not significant

Once measurement invariance was confirmed for the measurement model and the structural model tested, multi-group analyses were used for hypothesis testing. Results are summarized in Figure 2.

Figure 2. Structural model results for potential rapid and in-waiting entrepreneurs

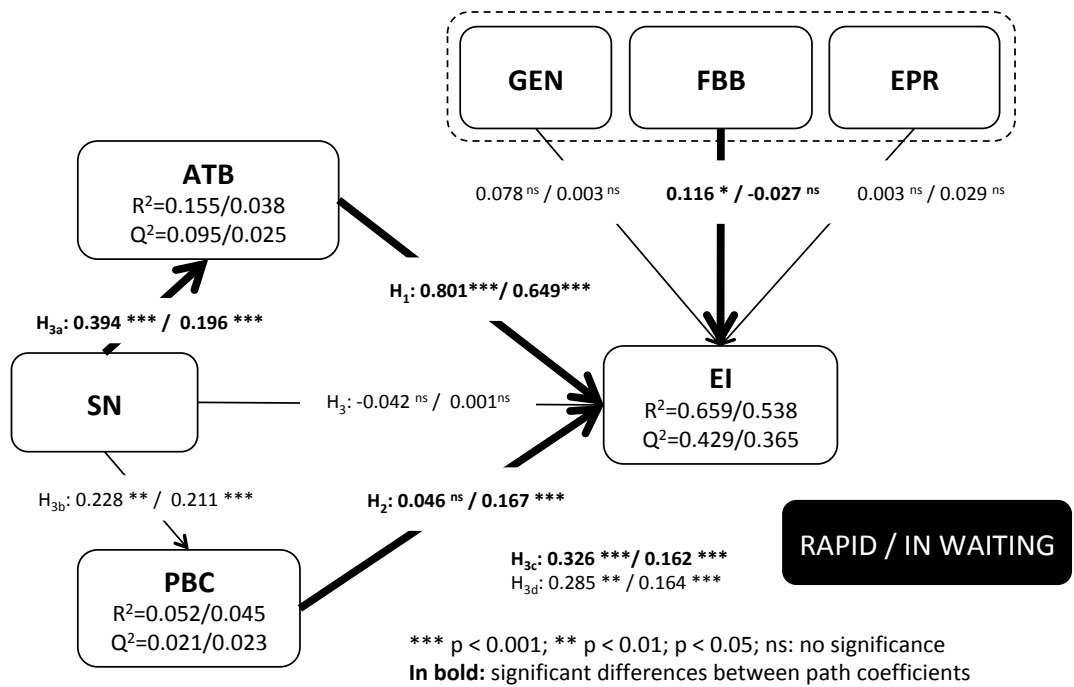

In this way, to test whether there were statistically significant differences in path coefficients between the two sub-samples, a parametric and three nonparametric tests were performed. The first was a parametric significance test for the difference of group-specific PLS-SEM results that assumes equal variances across groups (Sarstedt et al., 2011). When unequal variance across groups is assumed, the Welch-Satterthwait test can be applied (Sarstedt et al., 2011). Both tests were used for comparison, obtaining agreement in all hypotheses except for H3a and H3c (Table 6).

Subsequently, non-parametric approaches were applied. First, the Confidence Interval (bias corrected) approach was applied as suggested by Sarstedt et al. (2011). In this case, if the parameter estimate for a path relationship of one group does not fall within the corresponding confidence interval of another group and vice versa, there is no overlap and it can be assumed that the group-specific path coefficients are significantly different regarding a significance level α (Sarstedt et al., 2011). Second, the permutation-based procedure was applied, as suggested by Dibbern and Chin (2005) and Chin and Dibbern (2010). This technique is based on an approximate randomization test –where a subset of all possible data permutations between groups is made. Again, the level of agreement between data and all the hypotheses was good except for H3a and H3c (Table 6).

Finally, Henseler's PLS multi-group analysis was performed (PLS-MGA). A result was considered significant at 5% probability of error level, if the *p*-value was smaller than 0.05 or greater than 0.95 for a certain difference in group-specific path coefficients. The PLS-MGA method, as implemented in SmartPLS, is an extension of the original nonparametric Henseler's MGA method (as described, for example, by Sarstedt et al., 2011). As it can be seen in column $P_{PLS-MGA}$, statistical support was found for all the hypotheses except for H3, H3b and H3d.

Table 6. T-tests for multi-group analysis

| Path Coefficients | Hia | Coefficients-diff ( \| PRE - EIW \|) | tParam(EV) | | tparam(NEV) | | Bias Corrected 95% confidence intervals | | | | Pperm | | pPLS-MGA | | SUPPPORT |
|---|---|---|---|---|---|---|---|---|---|---|---|---|---|---|---|
| | | | | | | | EIW | | PRE | | | | | | |
| ATB ->EI | H1: (PRE>EIW) | 0.152 | 2.325 | * | 2.384 | * | 0.610 | 0.691 | 0.709 | 0.900 | 0.006 | ** | 0.015 | * | YES |
| PBC ->EI | H2: (EIW>PRE) | 0.121 | 1.567 | n.s. | 1.819 | n.s. | 0.120 | 0.216 | -0.042 | 0.153 | 0.062 | n.s. | 0.967 | * | YES |
| SN ->EI | H3: (PRE>EIW) | 0.043 | 0.580 | n.s. | 0.623 | n.s. | -0.045 | 0.048 | -0.152 | 0.054 | 0.274 | n.s. | 0.729 | n.s. | NO |
| SN -> ATB | H3a: (PRE>EIW) | 0.198 | 2.034 | * | 1.936 | n.s. | 0.150 | 0.269 | 0.262 | 0.568 | 0.030 | * | 0.031 | * | YES |
| SN ->PBC | H3b: (PRE>EIW) | 0.017 | 0.183 | n.s. | 0.197 | n.s. | 0.166 | 0.283 | 0.180 | 0.422 | 0.531 | n.s. | 0.407 | n.s. | NO |
| EPR ->EI | control: (PRE>EIW) | 0.027 | 0.397 | n.s. | 0.419 | n.s. | -0.013 | 0.072 | -0.104 | 0.090 | 0.342 | n.s. | 0.666 | n.s. | |
| FBB ->EI | control: (PRE>EIW) | 0.143 | 2.114 | * | 2.019 | * | -0.066 | 0.017 | 0.020 | 0.237 | 0.020 | * | 0.018 | * | |
| GEN ->EI | control: (PRE>EIW) | 0.074 | 1.105 | n.s. | 1.284 | n.s. | -0.038 | 0.047 | -0.007 | 0.166 | 0.140 | n.s. | 0.098 | n.s. | |
| **Indirect Effects** | | | | | | | | | | | | | | | |
| SN -> ATB, PBC ->EI | H3c: (PRE>EIW) | 0.164 | 2.239 | * | 1.787 | n.s. | 0.132 | 0.217 | 0.208 | 0.503 | 0.023 | * | 0.037 | * | YES |
| **Total Effects** | | | | | | | | | | | | | | | |
| SN ->EI | H3d: (PRE>EIW) | 0.121 | 1.314 | n.s. | 1.119 | n.s. | 0.121 | 0.232 | 0.139 | 0.474 | 0.134 | n.s. | 0.136 | n.s. | NO |

***p<0.001, **p<0.01, *p<0.05, ns = not significant (based on a Student t (4999) distribution with one tails)

In summary, these results provide empirical evidence that significant differences exist between potential rapid entrepreneurs and entrepreneurs-in-waiting. First, the impact of ATB on EI was significantly greater in potential rapid entrepreneurs than in "wait-and-see" entrepreneurs. Conversely, the impact of PBC on EI was significantly greater in entrepreneurs-in-waiting than in rapid entrepreneurs. On the other hand, SN was found to play an important role in the formation of EI. The impact of SN on ATB was significantly higher in rapid than among "in-wait" entrepreneurs; however, the impact of SN on PBC was similar in both groups. In agreement with previous studies, empirical evidence was obtained that SN has not a direct but an indirect and statistically significant effect on EI. This indirect effect was greater in potential rapid entrepreneurs than in entrepreneurs-in-waiting.

**Conclusion**

The purpose of this study was to investigate whether intended timing plays a moderating role in the relationships among the variables of the theoretical model of TPB. This paper examines differences in the impact of ATB, PBC and SN on EI of two groups of final-year university students who intended to become self-employed in the short or long term (PRE vs. EIW). In contrast with previous studies, the intended timing of entrepreneurship was considered in this study, and its moderating role in the model was assessed. This study provides empirical evidence that significant differences exist in some relationships of the variables of Ajzen's model between the two groups analysed.

In agreement with previous studies, this study reinforces evidence that a positive relationship exists between ATB and EI, and between PBC and EI. Nevertheless, according to the previous literature, the findings indicate that SN has not direct effects on EI (Krueger et al., 2000), but on ATB and PBC, which in turn have indirect effects on EI, as stated by Liñán and Chen (2009). Additionally, this study provides empirical evidence that the main predictor of EI is ATB rather than PBC or SN, as demonstrated by the analysis of explained variance of the EI construct (see Table 4). This reinforces the belief that entrepreneurship is more a matter of one's attitude (as stated by Krueger et al., 2000) than of individual's skills or support from significant others.

Regarding the moderating role of intended timing on the relationships studied, the results also show that there are some significant differences between PRE and EIW. First, the impact of ATB on EI is significantly greater in PRE than in EIW. This demonstrates that PRE are vocational entrepreneurs and their EI essentially depends on their ATB. Therefore, PRE are strongly motivated to start a business even although they lack resources, abilities or social support. Second, although the impact of PBC on EI was low in both groups, it was significantly greater in EIW than in PRE. This might contribute to their delay in going into business since EIW may feel they need more training and experience (e.g. by working as salaried employees) to improve their level of PBC.

In agreement with other authors (Autio et al., 2001; Krueger et al., 2000; Liñán and Chen,

2009), the empirical evidence obtained suggests that SN has no direct effect on EI in either group, but it has an indirect and statistically significant effect on EI mediated by ATB and PBC. The impact of SN on PBC was similar in both groups; however, the impact of SN on ATB was significantly higher in PRE than in EIW. A possible explanation to this may be that support from significant others in PRE's may reinforce PRE's entrepreneurial attitude to a greater extent than in EIW because the immediacy of entrepreneurial behaviour in the former requires an actual and stronger commitment than in the latter.

The relationship between education and entrepreneurial activity has been extensively examined (Bae et al., 2014; Espiritu-Olmos, and Sastre-Castillo, 2015). The results obtained in this study may have practical implications for entrepreneurship education and they can help policy makers develop more effective policies and programs to promote entrepreneurship.

In the light of the differences observed between rapid and wait-and-see entrepreneurs, universities are recommended to identify the profile of students with entrepreneurial intentions. Regarding PRE, entrepreneurship education programs are not as important as actions aimed at improving students' attitude toward entrepreneurship. Such actions can include granting awards for entrepreneurship projects, disseminating the history of successful entrepreneurs, creating entrepreneurship incubation centres, providing entrepreneurship counselling, and facilitating access to funding (seed capital, venture capital, etc.).

On the evidence of the relationship observed between PBC and EI, it is recommended that strategies addressed at potential entrepreneurs who set their time horizons in the long term (i.e. EIW) are focused on improving their entrepreneurial self-efficacy and PBC. Such strategies can include entrepreneurship education, corporate internships, project implementation, or the creation of partnerships with other entrepreneurs to build trusts.

Corporate internship programs can be especially effective in promoting EI among EIW. These programs can improve PBC and accelerate the process of gaining knowledge and experience. Also, internship programs can help students find business opportunities through the identification of tendencies, challenges to corporations and market gaps.

Another general recommendation to universities is to maintain contact with alumni, especially with those considered EIW, even years after their graduation. To such purpose, entrepreneurship promotion programs should include actions such as the formation of alumni associations, offering life-long training to alumni to improve or update their skills, and inviting them to participate or attend academic events.

Finally, entrepreneurial attitude should be inculcated to students through training strategies included in curricula, at the same time that entrepreneurship education is offered in postgraduate curricula. This way, graduate curricula would help PRE and EIW develop EI by improving their ATB, whereas postgraduate curricula would encourage EI in EIW, who need that their PBC is reinforced.

This study has some limitations. First, the cross-sectional design applied did not allow us to test causal relationships. Second, the use of self-report measures can be a limitation for the correct measurement of the constructs, and common-method variance is probably a shortcoming. Third, the theoretical model does not consider factors related to university environment such as university culture and climate, which can stimulate and/or hinder EI. Therefore, an institutional approach should also have been adopted, following the recommendations of Fuentelsaz et al. (2015). Fourth, the sample included university students in Spain. A larger sample including other cultures and countries would have provided more global and generalizable results. Finally, this study does not take into account other important variables such as entrepreneurial skills regarding innovation, risk-taking propensity or academic performance, which would help better explain the formation of EI. Some of these limitations have no solution and are present in other non-experimental disciplines. However, other limitations could be overcome so that future research confirms the validity of the conclusions and practical implications of this study.

The results of this study could also be complemented by future research focused on: (1) a longitudinal study in PRE and EIW first-year and last-year college students to explore the formation of EI over time; (2) identifying the year where training in ATB is more effective; (3) analysing university, family, social, geographic, socio-economic and environmental factors; (4) analysing EI using the TPB model and considering the branch of knowledge studied, since the prevalence of EI is not the same across studies; (5) investigating whether academic performance is determinant in the EI of college students.

**Acknowledgements**

This publication and research has been partially granted by INDESS (Research University Institute for Sustainable Social Development), Universidad de Cádiz, Spain.